\documentclass[aps,prb,superscriptaddress,showpacs,papersize=a4paper,twocolumn,amsmath]
{revtex4-1}%
\usepackage[pdftex]{graphicx}
\usepackage{subcaption}
\usepackage{etoolbox}
\usepackage[usenames,dvipsnames]{color}
\usepackage{mathtools}
\usepackage{amsmath}%
\usepackage{amsfonts}%
\usepackage{amssymb}

\begin{document}

\title{Conductivity and thermoelectric coefficients of doped SrTiO$_3$ at high temperatures}

\author{Kh.~G.~Nazaryan}
\affiliation{Moscow Institute of Physics and Technology, Dolgoprudny, Moscow Region, Russia}
\author{M.~V.~Feigel'man}
\affiliation{L.D.Landau Institute for Theoretical Physics, Chernogolovka, Moscow Region, 143432, Russia}
\affiliation{Moscow Institute of Physics and Technology, Dolgoprudny, Moscow Region, Russia}

\begin{abstract}
We developed a theory of electric and thermoelectric conductivity of lightly doped SrTiO$_3$ in the non-degenerate
region $k_B T \geq E_F$, assuming that the major source of electron scattering is their interaction with soft 
transverse optical phonons present due to proximity to ferroelectric transition. We have used kinetic equation approach
within relaxation-time approximation and we have determined energy-dependent transport relaxation time $\tau(E)$ by 
the iterative procedure. Using electron effective mass $m$ and electron-transverse phonon coupling constant $\lambda$
as two fitting  parameters, we are able to describe quantitatively a large set of the measured temperature dependences of 
resistivity $R(T)$ and Seebeck coefficient $\mathcal{S}(T)$ for a broad  range of electron densities studied experimentally
in recent paper~\cite{Behnia2020}.  In addition, we calculated Nernst ratio $\nu=N/B$ in the linear approximation over weak magnetic
 field in the same temperature range.

\end{abstract}

\date{\today}

\maketitle

\section{\label{intro} Introduction}

Very dilute 3D metal originating from band insulator Strontium Titanate (STO) due to tiny doping
($10^{-6}-10^{-3}$ conduction electrons per unit cell) demonstrates a number of rather unusual
properties~\cite{Spinelli2010,review1,review2}.  Their major common origin is the close proximity of 
insulating STO to a ferroelectric
transition, which lead to a giant low-temperature dielectric constant $\epsilon_0 \approx 20 000$.
In result, Coulomb interaction between conduction electrons is nearly vanishing, and  standard
phenomenology developed in the theory of normal metals is not applicable. Very low attainable density  $n$ 
of conduction electrons and its variability allow to study transport properties of doped STO in various
temperature regimes, from strongly degenerate Fermi gas at $k_B T \ll E_F$ to highly non-degenerate 
one at $k_B T \gg E_F$, with $E_F$ being the Fermi energy.  The latter range is in the focus of recent experimental
studies~\cite{Behnia2020,Behnia2017}; see also similar mobility data in Ref.~\cite{Verma} and thermoelectric data
in Ref.~\cite{Cain}.
 It was found~\cite{Behnia2017} that at high temperatures $T \geq 300$ K
conductivity drops below Mott-Ioffe-Regel limit and, moreover, relaxation  rate $1/\tau$ becomes larger than its
apparent quantum limit $k_B T/\hbar$. Later on, the paper~\cite{Behnia2020} demonstrates that the account for the
temperature-dependent renormalization of the effective mass $m(T)$ makes the above contradiction less drastic.
However, the behavior of $m(T)$ found in Ref.~\cite{Behnia2020} via the fitting of their data for resistance $R(T)$
and Seebeck coefficient $\mathcal{S}(T)$  is somewhat unexpected. Namely, mass renormalization $m(T)/m_0$ 
fitted in~\cite{Behnia2020}
depends not only  on temperature but also on the electron density $n$, which should not be the case in the non-degenerate 
region $k_BT \gg E_F$ studied. Also, initial increase on $m(T)$ with $T$ growth is replaced by
its drop with temperature at $ T $ above 300 K, which does not have physical 
explanation.

In the present paper we reconsider theoretically the issue of electric conductivity and thermoelectric response 
in a non-degenerate electron gas, interacting with soft transverse optical (TO) phonons.  These type of phonons exist in STO due to
its proximity to ferroelectric transition~\cite{YamadaShirane,review1,review2,Behnia2020}.  
Scattering of non-degenerate electrons due to bi-quadratic coupling to transverse optical phonons was considered in 
somewhat different context in Refs.~\cite{Levanyuk1,Levanyuk2}; see also older experiments~\cite{old1,old2}.
Recent paper~\cite{Yudson21} provided another approach to kinetic properties based upon the idea of
dominant role of electron scattering on two TO phonons; we will compare our results with those of Ref.~\cite{Yudson21}
in the final part of the paper.

 We use kinetic equation approach
in the form close to the one discussed in Ref.~\cite{Theory} and express both resistivity $R(T)$ and Seebeck
coefficient $\mathcal{S}(T)$ in terms of the energy-dependence transport scattering time $\tau(E,T)$.  We emphasize  that
the energy-dependence of $\tau(E,T)$ is not weak, and this is the reason why simple proportionality
relation~\cite{Behnia2020}  between Seebeck coefficient $\mathcal{S}(T)$ and thermodynamic  
entropy per electron $S(T)$ is not valid actually.  Indeed, $\mathcal{S}(T) = \frac{k_B}{e} S(T)$ follows once
$\tau(E,T)$ can be replaced by energy-independent $\tau(T)$, as we will show  below.

Real band structure of SrTiO$_3$ is rather complicated (see for example~\cite{Dirk2011}), 
and detailed calculation of transport coefficients is hardly possible without use of heavy numerical 
procedures based upon band-structure calculations. 
Some examples  of the latter type of approach can be found in Refs.~\cite{numeric1,numeric2}.
While Ref.~\cite{numeric1} reported good agreement of computed high-$T$ mobility with experimental data,
the later Ref.~\citep{numeric2} asserts that numerically exact band-structure calculation does over-estimate 
mobility at high $T$ by a factor close to 10, while providing correct $T$-dependence. Paper~\citep{numeric2} hints
that the discrepancy originates from temperature-dependent polaron effect leading to the increase of effective electron
mass with temperature. Later paper~\citep{numeric3} presents the results obtained by numerical implementation of
this type of approach: strong polaron effects due to electron-phonon interactions were taken into account and
good agreement with numerically obtained data on mobility was obtained.
 Moreover it was found that strong incoherent effects (demonstrated by
broad electron spectral function) are developing at $T \sim 250-300^\circ$K.

 We will accept general idea of importance of $T$-dependent mass renormalization and
 develop a semi-quantitative theory based upon the simplest model of parabolic electronic spectrum 
$E(p)=p^2/2m$ with temperature-dependent effective mass $m=m(T)$. We believe that proximity of STO
to ferroelectric transition is the key feature of this material; most probably it is responsible
for its anomalous properties, thus we feel it useful to consider the effects of electron interaction with TO phonons
in somewhat more details. Thus our approach is clearly alternative to the one developed in Refs.~\cite{numeric1,numeric2}
where coupling to  high-energy longitudinal phonons (LO) was studied as a major source of electron scattering. 
We emphasize, to avoid any confusion,
 that in the whole temperature we consider, the temperature-dependent TO phonon gap  $\hbar \omega_T(T) \ll k_B T$, 
this condition is well fulfilled up to $1000^{\circ}$ K, as  Fig.6 of Ref.~\cite{Behnia2020} demonstrates.
At the same time, LO phonons have much higher energy gaps.

 Our goal is to provide simple model of
electron scattering leading to reasonable agreement with the data in the high-temperature region $T \gg E_F(n)$
for both conductivity $\sigma(T)$, Seebeck coefficient $\mathcal{S}(T)$ and Nernst coefficient $\nu(T)$.
We will show that straightforward kinetic equation approach leads (once
 moderate  mass renormalization, weakly $T$-dependent at high temperatures, is taken into account) to
 good agreement with the data for $\sigma(T)$ and $\mathcal{S}(T)$ in a broad range of temperatures 
$100 < T < 700^{\circ}$K and for a broad range of electron densities $1.4\cdot 10^{17} \leq n \leq 3.5 \cdot 10^{20} $cm$^{-3} $.

The rest of the paper is organized as follows:  we formulate our model in Sec.~\ref{formulation}, then in Sec.~\ref{kineq}  we derive Boltzmann 
kinetic equation for electrons interacting quadratically with TO phonons, and expose the major points of its solution
in terms of the effective relaxation time $\tau_{TO}(p)$.  Next, in Sec.~\ref{transport} we discuss calculation of electric conductivity
and thermoelectric coefficients in terms of this effective relaxation time (Subsec.~\ref{condSeeb} and \ref{nernst}) and then present
our final results and compare them with experimental data in Sec.~\ref{results}.  Sec.~\ref{conclusions} contains our conclusions.



\section{\label{formulation} Formulation of the model}

We are interested in the high-temperature region where quantum statistics of electrons is irrelevant
and it is sufficient to consider each electron individually interacting with lattice phonons.  The
Hamiltonian of an electron coupled to transverse optical phonons is
\small
\begin{equation}
H =  \int d^3r \left[ -\hbar^2\frac{\psi^\dagger(\mathbf{r})\nabla^2\psi(\mathbf{r}) }{2m(T)}
 + 
   g \rho_m \, \psi^\dagger(\mathbf{r}) \psi(\mathbf{r}) \mathbf{u}^2(\mathbf{r}) \right]
\label{H1}
\end{equation}
\normalsize where $\rho_m=5.11$ g/cm$^3$ is the STO mass density,  $\psi(\mathbf{r})$ is the electron annihilation operator and $\mathbf{u}(\mathbf{r})$ is the atomic 
displacement field related to the optical transverse mode (TO), $\nabla \mathbf{u}=0$.  Dispersion of this mode
is $\omega(k) = \sqrt{\omega_{TO}^2 + (sk)^2}$  where  $s \approx 7.5\cdot 10^5$ cm/s is the TO mode velocity~\cite{Svel}
 and
$\omega_{TO}$ is the temperature-dependent gap which grows with temperature~\cite{Behnia2020,YamadaShirane,Baurle};
in particular, $\omega_{TO}$ varies between 5 meV and 18 meV when $T$ grows from 100 K to 800 K.
Bilinear coupling (\ref{H1}) to transverse phonons was proposed recently as a possible mechanism for superconductivity 
in doped STO~\cite{Dirk2019}; we develop this idea further in a separate publication~\cite{KiselevFeigelman}.
Another possible type of  electron's coupling  to transverse  TO phonons may contain vector term of the type of
$i(\psi^+\nabla\psi - h.c.) \mathbf{u}$, but we will not consider its effects here.

Effective  band mass of electrons depends~\cite{Behnia2014} on their concentration at low temperatures in the degenerate region
$T \ll E_F$; in particular, for zero-temperature mass $m(0)/m_0 \approx 1.8$ for lowest concentrations $n \leq 10^{18}$ cm$^{-3}$,
while at higher doping $n \geq 4\cdot 10^{18}$ cm$^{-3}$ the same ratio $m(0)/m_0 \approx 4$; here $m_0$ is the free electron mass.
 However, at high temperatures
$T\gtrsim 150$ K Hall mobility is $n$-independent, as follows from the data present in Ref.~\cite{Behnia2020}.  The same is
expected for the effective mass, which may however depend on temperature, for two different reasons. First one is due to
 non-parabolic dispersion with decreasing  effective curvature of $\xi(\mathbf{p})$ at large momenta $|\mathbf{p}|$, see Fig.~3 in 
Ref.~\cite{Dirk2011}: at high $T$ carriers with energies much above the bottom of the band are most relevant, and effective
mass seen in transport characteristics increases;  moreover, at high temperatures the major contribution to transport is 
provided by the lowest band since at large $|\mathbf{p}|$ it provides the largest density of states. The second reason for the mass
enhancement is the polaron effect due to coupling with TO phonons~\cite{numeric3}. 
 A special remark is in order concerning our use of parabolic model of spectrum in the situation of 
highly anisotropic band~\cite{Dirk2011}.  The point is that at high temperatures we consider, the 
scattering rate $1/\tau_{TO}$ is rather high so electrons quickly forget the directions of their momenta,
thus isotropic approximation looks acceptable, at least for the beginning.
Below we will treat  $m(T)$ as a phenomenological fitting function, 
common for all concentrations $n$. 

The coupling constant $g$ in Eq.(\ref{H1}) has a natural dimension $\textrm{Length}^3/\textrm{Time}^2$.  We present it
in the form
\begin{equation}
g = \lambda {a^3} \omega_L^2
\label{lambda}
\end{equation}
where $a=0.39$ nm is the STO lattice constant, $\hbar\omega_L \approx 100$meV is the highest energy gap of all
longitudinal optical modes and
$\lambda$ is some dimensionless constant, to be determined below from the fit of our theory  to the data.

\begin{figure*}[t]
\begin{centering}
\begin{subfigure}{.99 \linewidth} \begin{centering}
\includegraphics[width=1\linewidth]{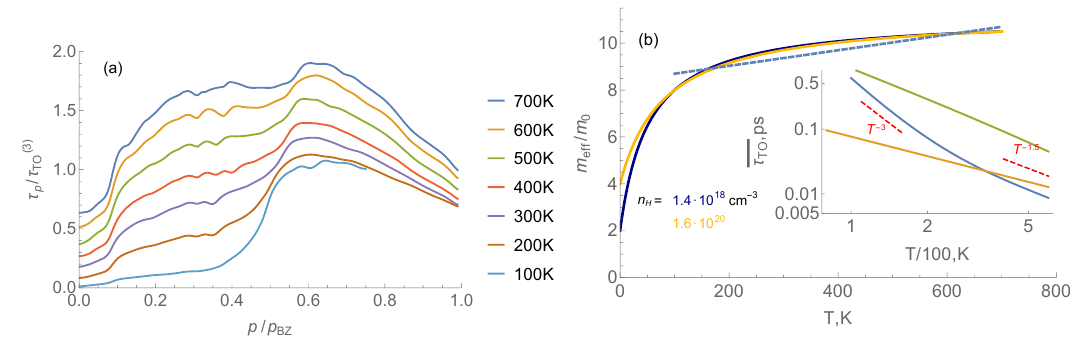} 
\par\end{centering}\label{fig:1m}  \end{subfigure}\hfill
\par\end{centering} 
\caption{\textbf{a)} Lines represent the results of the $3^{\text{rd}}$ iteration of the method of successive approximations for the inverse scattering time normalized to the inverse Planckian time $\tau_p/{\tau_{TO}^{(3)}}$.
Here, $p_{BZ}=\hbar \pi /a$ stands for the first Brillouin zone and $\tau_p=\hbar/{k_B T}$. \textbf{b)} Full lines show the expected qualitative temperature dependency of the effective mass. The dashed line stands for the linear dependency chosen in the present paper. The inset shows the comparison of the scattering time including 2TO phonons averaged as $\overline{\tau_{TO}}=\langle \tau_{TO} \rangle/\langle 1 \rangle$ (blue line), AC phonon (green line) and the Planckian time (orange line) \label{fig:1}} 
\end{figure*} 


\section{\label{kineq} The kinetic equation and  effective relaxation time}

In this Section we consider soft TO phonons as the major source of electron scattering.  In addition we consider the role of
usual acoustic phonons and demonstrate that their contribution to the scattering rate is much smaller and can be neglected
in the first approximation.

\subsection{\label{TOphonons} Transverse optical phonons}

We will study electric and thermoelectric transport in an electron system
using the Boltzmann kinetic equation. To find the transport coefficients within
linear response theory, we expand  the distribution function  
as $f_{\boldsymbol{p}}\approx n_{p}+\delta n_{\boldsymbol{p}}$.
Here, $n_{p}=\left[\exp\left(\beta\xi_{p}\right)+1\right]^{-1}$ is
the Fermi-Dirac distribution with $\beta=1/T$, and $\xi_{p}=E(p)-\mu$
with the chemical potential $\mu$. This helps to write the linearized
Boltzmann equation in presence of both electric field and temperature gradient:
\begin{align}
 & \left(-e\mathbf{E}-\xi_{\mathbf{p}}\frac{\nabla_{\mathbf{r}}T}{T}\right)\cdot v_{\mathbf{p}}\frac{\partial n_{p}\left(\xi_{\mathbf{p}}\right)}{\partial\xi_{\mathbf{p}}}=I_{TO}\{\delta n_{\boldsymbol{p}}\}
\label{Beq}
\end{align}
where   $\mathbf{v}_{\mathbf{p}} = \partial\xi_{\mathbf{p}}/\partial \mathbf{p} = \mathbf{p}/m^* $  is the group velocity. 
An exact expression for the collision integral $I_{TO}\{\delta n_{\boldsymbol{p}}\} $ which describes
electron scattering by two TO phonons is of the following form (for the derivation, see Appendix~\ref{app:collint}):
\begin{widetext}
\small
\begin{align}
 & \frac{1}{A}I_{TO}\{\delta n_{\boldsymbol{p}}\}=
\label{ITO}\\
 & =2\int_{\boldsymbol{k_{1}k_{2}}}\frac{\delta\left(E_{p'}+E_{k_{1}}-E_{p}-E_{k_{2}}\right)}{\omega(k_{1})\omega(k_{2})}\left(\delta n_{\boldsymbol{p}}\left(\left(N_{k_{2}}-N_{k_{1}}\right)n_{p'}-\left(N_{k_{1}}+1\right)N_{k_{2}}\right)+\delta n_{\boldsymbol{p'}}\left(n_{p}\left(N_{k_{2}}-N_{k_{1}}\right)+\left(N_{k_{2}}+1\right)N_{k_{1}}\right)\right)\nonumber \\
 & +\int_{\boldsymbol{k_{1}k_{2}}}\frac{\delta\left(E_{p'}-E_{k_{1}}-E_{p}-E_{k_{2}}\right)}{\omega(k_{1})\omega(k_{2})}\left(-\delta n_{\boldsymbol{p}}\left(N_{k_{1}}N_{k_{2}}+\left(1+N_{k_{2}}+N_{k_{1}}\right)n_{p'}\right)+\delta n_{\boldsymbol{p'}}\left(\left(N_{k_{2}}+1\right)\left(N_{k_{1}}+1\right)-n_{p}\left(1+N_{k_{2}}+N_{k_{1}}\right)\right)\right)\nonumber \\
 & +\int_{\boldsymbol{k_{1}k_{2}}}\frac{\delta\left(E_{p'}+E_{k_{1}}-E_{p}+E_{k_{2}}\right)}{\omega(k_{1})\omega(k_{2})}\left(\delta n_{\boldsymbol{p}}\left(\left(1+N_{k_{1}}+N_{k_{2}}\right)n_{p'}-\left(N_{k_{1}}+1\right)\left(N_{k_{2}}+1\right)\right)+\delta n_{\boldsymbol{p'}}\left(n_{p}\left(1+N_{k_{2}}+N_{k_{1}}\right)+N_{k_{2}}N_{k_{1}}\right)\right)\nonumber 
\end{align}
\end{widetext}
\normalsize
where 
$N_{k}=\left[\exp\left(\beta\omega_{k}\right)-1\right]^{-1}$
is the Bose-Einstein distribution for TO phonons, and $A=\frac{\pi}{2}{g^{2}}$.
 We use the short-cut notation for the integrals $\int_{\boldsymbol{k_{1}k_{2}}}=\int_{BZ}\frac{d^{3}\boldsymbol{k_{1}}d^{3}\boldsymbol{k_{2}}}{\left(2\pi\right)^{6}}$,
where BZ stands for the first Brillouin zone (we take it in spherical approximation). 
Integration over momenta in Eq.(\ref{ITO}) is carried out under the condition of conservation of the total momentum
(we neglect Umpklapp processes).
Conservation of the total energy is expressed in the explicit form in Eq.(\ref{ITO}) by the delta-functions; the forms
of their arguments demonstrate that the 1st line corresponds to the scattering of an electron by the TO phonon, 
while 2nd and 3rd lines correspond to emission and absorption of  two TO phonons, correspondingly.

Scattering processes change electron energy by the typical amount $\sim k_B T$ that is of the order of typical electron
energy in the non-degenerate region $k_B T \gg E_F$ we consider here.  Therefore the only way to proceed is to solve the
kinetic equation (\ref{Beq},\ref{ITO}) numerically and to find non-equilibrium correction to the distribution function
$\delta n_{\boldsymbol{p}}$ originating due to electric field $\mathbf{E}$ or temperature gradient $\nabla T$.
It is useful to present the result of such a computation in terms of some \textit{effective} relaxation time
$\tau_{TO}(p,T)$ according to the definition below:
\begin{align}
 & I_{TO}\{\delta n_{\boldsymbol{p}}\} \approx -\frac{\delta n_{\boldsymbol{p}}}{\tau_{TO}(p)}
\label{RTA}
\end{align}
Effective relaxation rate $\tau^{-1}_{TO}(p;T)$ which provides the best approximation of the exact
$ I_{TO}\{\delta n_{\boldsymbol{p}}\}$ by the form of Eq.(\ref{RTA}), can be found by means of the successive approximation
 method, described in details in the Appendix~\ref{app:calc}.

Briefly, it works as follows.
First we note that in the linear regime over weak electric field $E$, modification of the electron
 distribution function can be written in the form (which follows by comparison between Eq.(\ref{Beq}) and Eq.(\ref{RTA})):
\begin{align}
\label{noGrad}
 & \delta n_{\boldsymbol{p}}\approx e\,\tau_{TO}(p)\cdot\left(\boldsymbol{v_{p}},\mathbf{E}\right)\frac{\partial n_{F}\left(\xi_{p}\right)}{\partial\xi_{p}}
\end{align}
The form of Eq.(\ref{noGrad}) assumes it is possible to neglect temperature gradient $\boldsymbol{\nabla}_{r}T$ 
while calculating   $\delta n_{\boldsymbol{p}} $ and $\tau_{TO}(p)$, and we check it \textit{a posteriori},
see the end of Appendix~\ref{app:Results}.
The only unknown function here is $\tau_{TO}=\tau_{TO}(|\boldsymbol{p}|)$. 
 In the case of degenerate Fermi-gas relevant momentum $p=p_F$ and we are left with just a 
single number $\tau$ for relaxation time.  At high temperatures $ T \geq E_F$ the situation is more complicated.
Relaxation time $\tau_{TO}(|\boldsymbol{p}|)$ appears to be a strong function of the electron energy.
To find this function we start from the simplest trial with $\tau_{TO} (p) = \tau_{TO}$ independent of energy, and
substitute the corresponding $ \delta n_{\boldsymbol{p}}^{(0)} $ into the R.H.S. of Eq.(\ref{ITO}).  The result is then
represented in the form of Eq.(\ref{noGrad}) with somewhat different $\tau_{TO}^{(1)}(p) $ which is now energy-dependent.
Then the process is repeated upon convergence (see Appendix~\ref{app:Results}).  We have found that the difference of the obtained
$\tau_{TO}^{(2)}(p) $ and  $\tau_{TO}^{(3)}(p) $ is already quite small, thus it is possible to use $\tau_{TO}^{(3)}(p) $
as our final approximation for the relaxation rate.  It is shown in Fig.~\ref{fig:1}a for a number of different temperatures.
One can clearly see in Fig.~\ref{fig:1}a that the momentum dependence of scattering time $\tau_{TO}(p)$ is not weak. 
This is why the Seebeck coefficient $\mathcal{S}(T)$ in not just proportional to the thermodynamic entropy per electron $S(T)$ 
as it was supposed in the paper~\citep{Behnia2020}. 
Instead, we will employ  an approach provided in Ref.~\cite{Theory} where various transport coefficients  were found
in terms of effective energy-dependent relaxation rate $1/\tau(p)$.
Note that our definition of effective relaxation time $\tau_{TO}(p) $ is not  \textit{a priori}  invariant with respect to
different sources of non-equilibrium; one could imagine that electric field $\mathbf{E}$ and temperature gradient $\nabla T$
could lead to the solutions of Boltzmann equation which correspond to different $\tau_{TO}(p) $ functions.
In fact, it is not the case, as we demonstrate in the Appendix C2,  the results for "purely electric" source and 
for the case of  zero-electric-current geometry, when electric field effect is completely compensated by the effect of thermal gradient, are very close to each other. It provides our effective relaxation time $\tau_{TO}(p) $ with more substantial physical meaning.

\subsection{\label{acoustic} Acoustic phonons}

The other mechanism that can probably contribute to the transport processes in the discussed temperature range is the scattering on the longitudinal acoustic (AC) phonons. The Hamiltonian of an electron coupling to an AC phonon reads
\begin{align}
 H_{ac}=\int D_{ac}\, \psi^\dagger(\mathbf{r}) \psi(\mathbf{r})\, \text{div}\, \mathbf{u}(\mathbf{r})
\end{align} 
where $\mathbf{u}(\mathbf{r})$ in this case is the atomic displacement field related to the acoustic mode. Their dispersion is given as $\omega (k)=v_s k$, with $v_s\approx 8.1\cdot 10^5$ cm/s the sound velocity in STO\cite{soundvel}. The constant $D_{ac}\approx 2.87$ eV is the deformation potential in STO\cite{defpot}. 
The interaction with the acoustic phonons results a scattering time which can be approximated via the following expression\cite{acoustic}
\begin{align}
\tau_{AC} = \frac{(8\pi)^{1/2}\hbar^4\rho_mv_s^2}{m^{3/2} D_{ac}^2 (k_B T)^{3/2}} 
 \end{align}
This result is compared to the contribution from the scattering time on 2TO phonons (see the inset on Fig.~\ref{fig:1}b)). In the most part of the studied region the scattering rate due to TO phonons $1/\tau_{TO}$ turns out to be considerably greater than the acoustic phonon's contribution  $1/\tau_{AC}$. Therefore, we may conclude that the consideration of only 2TO processes is an appropriate approximation, although for the temperatures at the lower end of the considered range the acoustic phonons can change the results slightly.

\section{\label{transport} Transport coefficients in terms of effective relaxation time}

\subsection{\label{condSeeb} Conductivity and Seebeck coefficient.}

Electric conductivity $\sigma$ and the Seebeck coefficient $\mathcal{S}$  can be expressed via $\tau(p)$
as follows~\citep{Theory}:
\begin{align}
 & \sigma=\frac{e^{2}}{m}\mathcal{N} \left\langle \tau_{TO}(p) \right\rangle \label{sigma}\\
 & \mathcal{S}=-\frac{1}{eT}\frac{ \left\langle \xi_{\boldsymbol{p}}\tau_{TO}(p) \right\rangle }{ \left\langle \tau_{TO}(p) \right\rangle }\label{S}
\end{align}
where  $ \left\langle \dots\right\rangle $
denotes the average $ \left\langle X_{\mathrm{p}} \right\rangle =-\frac{4}{3\mathcal{N}}\int_{\mathbf{p}}X_{\mathbf{p}}\left(\xi_{\mathbf{p}}+\mu\right)\frac{\partial n_{p}\left(\xi_{\mathbf{p}}\right)}{\partial\xi_{\mathbf{p}}}$
with $\mathcal{N}=2\int_{\boldsymbol{p}}n_{p}(\xi_{p})$.
While Eq.(\ref{sigma}) for conductivity is evident,  the result (\ref{S}) for the Seebeck coefficient needs some comments.
Indeed, if one assumes $\tau(p)$ to be some constant (independent on $p$), then Eq.(\ref{S}) leads to the relation
$\mathcal{S}(T) = \frac{k_B}{e} S(T)$ used for the analysis developed in Ref.~\cite{Behnia2020}. In fact, the $p$-dependence
of relaxation rate is very strong, as shown in Fig.~\ref{fig:1}a.

The equations (\eqref{sigma}, \eqref{S}) can be derived as follows. We allow for the presence of both electric field  $\mathbf{E}$
and temperature gradient $\nabla T$ and express $\delta n_\mathbf{p}$ using Eq. \eqref{Beq} for the collision integral in the RTA form \eqref{RTA}.  The electric and thermal currents are then determined as follows:
 \small
\begin{align}
\left(\begin{array}{c}
\mathbf{J}_{\mathbf{E}}\\
\mathbf{J}_{T}
\end{array}\right)&=2\int_{\mathbf{p}}\left(\begin{array}{c}
-e\\
\xi_{\mathbf{p}}
\end{array}\right)\boldsymbol{v_{p}}\delta n_{\mathbf{p}}=\left(\begin{array}{cc}
\sigma & \alpha\\
\beta & \kappa'
\end{array}\right)\left(\begin{array}{c}
\mathbf{E}\\
-\mathbf{\nabla}_{\mathbf{r}}T
\end{array}\right)
\end{align}
\normalsize
where $\sigma$ is electric conductivity, $\alpha={\beta}/{T}$ due to Onsager reciprocity relations and 
Seebeck coefficient $\mathcal{S}={\alpha}/{\sigma}$.  The use of $ \delta n_{\mathbf{p}}$ in the form of 
Eq.(\ref{Beq}) leads then to~\eqref{sigma}, \eqref{S}.  Note that now we consider kinetic equation in presence
of both electric field and temperature gradient; we checked in Appendix~\ref{app:Results} that the presence of $\nabla T$
 does not change the function $\tau_{TO}(p)$ found previously.

\begin{figure*}[ht!]
\begin{centering}
\begin{subfigure}{.49 \linewidth} {\begin{centering}
\includegraphics[width=1\columnwidth]{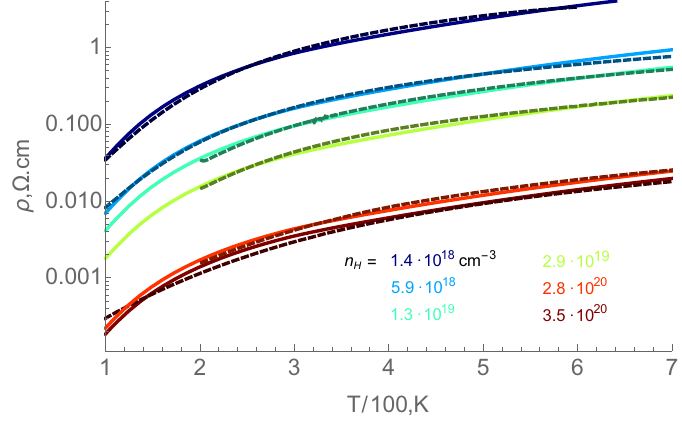} \caption{ Temperature dependency of the resistivity extended
to 700 K for several Nb doped $\text{SrTiO}_{3}$ crystals.}\label{fig:Rho}
\par\end{centering}}     \end{subfigure}
\begin{subfigure}{.49 \linewidth} {\begin{centering}
\includegraphics[width=1\columnwidth]{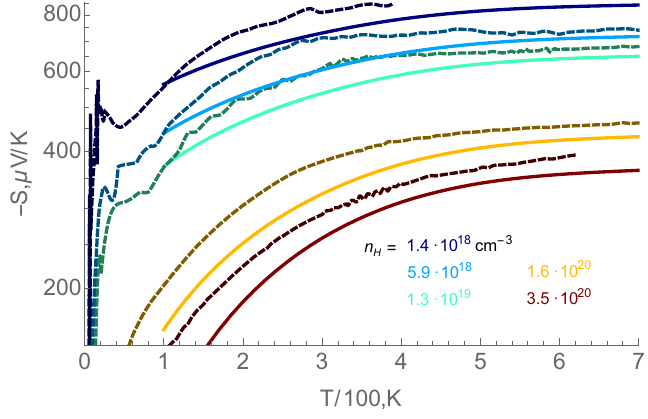}  \caption{The Seebeck coefficient as a function of temperature
for four Nb doped $\text{SrTiO}_{3}$ samples from 100 K to 700 K}\label{fig:Seeb}
\par\end{centering}} \end{subfigure}
\par\end{centering} 
\caption{Experimental data~\cite{Behnia2020} are shown by dashed lines, while full lines are used 
for the theoretical results.} 
\end{figure*} 

\begin{figure}[h]
{ \begin{centering}
\includegraphics[width=0.75\linewidth]{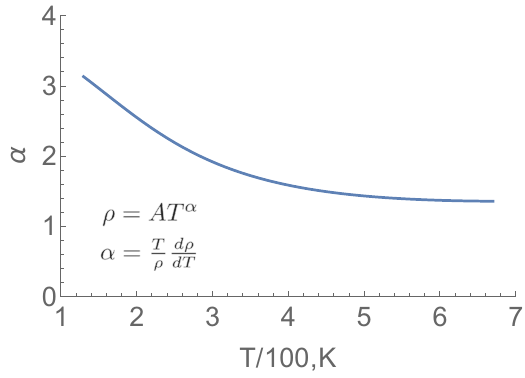} 
\par\end{centering}}
\caption{\label{alpha} Temperature dependency of the exponent $\alpha$ of
the resistivity: $\rho = A T^{\alpha}$}
\end{figure} 

\begin{figure}[h]
{ \begin{centering}
\includegraphics[width=1\linewidth]{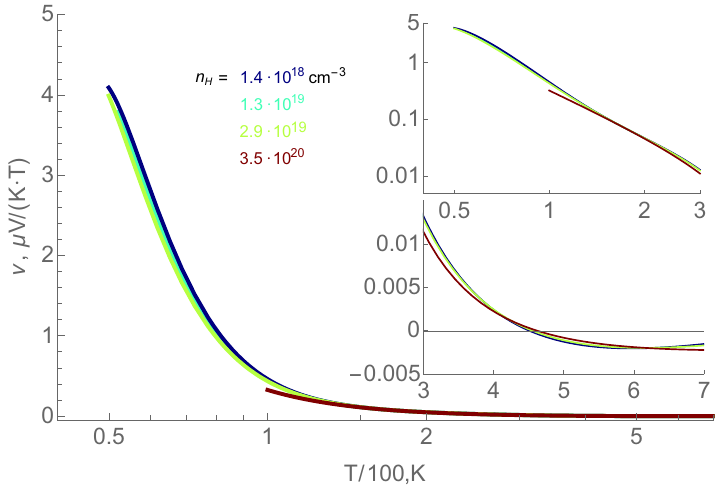} 
\par\end{centering}}
\caption{\label{fig:Nernst} Nernst coefficient $\nu(T)$ predicted for several Nb doped $\text{SrTiO}_{3}$ samples from 50 K to 700 K. The lower inset shows the temperature dependency of the Nernst coefficient in the range  300 - 700 K. 
The upper inset shows the same dependency from 50 K to 300 K.}
\end{figure} 

\subsection{\label{nernst} Nernst coefficient.}
In presence of magnetic field $\mathbf{B}$, an applied electric field $\mathbf{E} \perp \mathbf{B}$ leads to the temperature gradient
transverse to both $\mathbf{E}$ and $\mathbf{B}$. The corresponding response is called Nernst signal, $N = E_y/\nabla_x T$.
To calculate it we start from the general expression (See Sec.5 of the book~\cite{KamranBook}) for $N$ in terms of electric conductivity tensor 
$\sigma_{\alpha\beta}$ and thermoelectric tensor $\alpha_{\alpha\beta}$:
\begin{align}
 N=\frac{E_y}{\nabla_x T}=\frac{\alpha_{xy}\sigma_{xx}-\alpha_{xx}\sigma_{xy}}{\sigma_{xx}^2+\sigma_{xy}^2}
 \end{align} 
We concentrate here upon the limit of weak magnetic field and thus consider lowest-order terms in the magnitude of $B$:
\begin{align}
&\sigma_{xx}\approx\sigma;\,\,\; \alpha_{xx}\approx\alpha=-\frac{e\mathcal{N}}{mT} \left\langle \xi_p\tau_{TO}(p) \right\rangle;\\
&\sigma_{xy}\approx \frac{e^3 \mathcal{N}}{m^2}B\left\langle \tau^2_{TO}(p)\right\rangle;\,\,\; \alpha_{xy}\approx-\frac{e^2\mathcal{N}}{m^2T}B \left\langle \xi_p\tau^2_{TO}(p) \right\rangle 
\end{align}
These equations lead us to the expression of the linear Nernst coefficient $\nu=dN/dB|_{B=0}$:
 \begin{align}
\nu=\frac{\left\langle \tau^2_{TO}(p) \right\rangle  \left\langle \xi_{\textbf{p}}\tau_{TO}(p) \right\rangle-\left\langle \tau_{TO}(p)\right\rangle \left\langle \xi_{\textbf{p}}\tau^2_{TO}(p) \right\rangle}{T\left\langle \tau_{TO}(p)\right\rangle^2}
\label{nu}
\end{align}
Eq.(\ref{nu}) will be used below to compute Nernst coefficient $\nu(T)$ as function of temperature.


\subsection{\label{results} Results for the electric and thermoelectric coefficients.}

The obtained results for the scattering time $\tau_{TO}$ enabled calculation
of the electric and thermoelectric coefficients using \eqref{sigma}
and \eqref{S}.  The results of  numerical computation for the electric
resistivity and the Seebeck coefficient are provided on Fig.~\ref{fig:Rho}
and Fig.~\ref{fig:Seeb} respectively.  The mass $m(T)$ and the coupling constant $\lambda$ were fitted to 
experimental data from Ref.~\cite{Behnia2020} in the  way to minimize relative errors between the data and the theory.
We emphasize that the single set of $\lambda$ and $m(T)$ was used to fit the data for five different doping concentrations 
(apart from $n=5.9\cdot 10^{18}$ cm$^{-3}$), and relative error is about 7-9\%. Namely, we choose $\lambda = 0.88$ (with a relative error of 5\%) and
effective mass $m(T)$ which linearly interpolates  between $8.7 m_0$ for $100$ K to $10.7 m_0$ for $700$ K, 
see Fig.\ref{fig:1}b. For the concentration $n_H=5.9\cdot 10^{18}$ cm$^{-3}$ we found $\lambda\approx 0.78$, which is almost 10\% less than for the others. 


Dependence of resistivity on temperature $\rho(T)$ in a broad temperature range cannot be present by any single power law.
However, local "effective exponent"  $\alpha(T)$ can be introduced via the relation $\alpha(T)=\frac{T}{\rho } \frac{d\rho}{dT}$.
Our theoretic  result for $\alpha(T)$  corresponding to the full curves in Fig.~\ref{fig:Rho}  is present in  Fig.~\ref{alpha}.
Comparison with the date present in Fig.S1a of Ref.~\cite{Behnia2020} show excellent agreement for low electron concentrations;
in particular, peculiarity of large $\alpha \approx 3$ in the temperature range 100-200K  is clearly reproduced by our theory,
as well as the approach to lower value of $\alpha \approx \frac32$ at highest temperatures.
At higher electron concentrations $n \geq 10^{20}$ the data~\cite{Behnia2020}  does not show so large values of $\alpha(T)$; this is due to the fact the specific temperature range around 150 Kelvins lies below Fermi temperature $E_F(n)/k_B$, while our calculations were done for non-degenerate electrons only.

Our predictions for the Nernst coefficient $\nu(T)$ are shown in Fig.~\ref{fig:Nernst}. We present our results for different
electron densities but always for temperatures $T$ above the corresponding Fermi energy $E_F(n)$.
At these high temperatures $\nu(T)$ is independent on density of electrons, similar to the behavior of  mobility $\mu(T)$.
While at low temperatures Nernst signal is known to be strongly $n$-dependent~\cite{Behnia2013}, 
we are not aware of any measurement of this quantity in STO in the non-degenerate region $T \gg E_F$.
Upper inset to Fig.~\ref{fig:Nernst} shows $\nu(T)$ in double-logarithmic scale, where one can infer a power-law-like behavior 
$\nu(T) \propto T^{-n}$ with $n \approx 3.5$. Lower inset is concentrated in the high-$T$ region where sign change
 of $\nu(T)$ is predicted at $T\approx 450$ K.

\section{\label{conclusions}Conclusions}

We developed a theory of conductivity and thermoelectric effects in doped Strontium Titanate in the non-degenerate region
$T \gg E_F(n)$, assuming that the major source of electron scattering is due to processes which involve two soft optical (TO) phonons.
The strength of the interaction between electrons and these TO phonons is determined by single dimensionless parameter $\lambda$.
We have found good agreement between our theory and experiment~\cite{Behnia2020} for both conductivity and Seebeck coefficient and 
for (almost) all electron concentrations, using single value of  $\lambda \approx 0.88$ and an assumption of an enhanced effective mass, with the ratio $m^*(T)/m_0 $ varying between about 8.7 and 10.7 in the temperature interval 100-700 Kelvin 
(independent on $n$). 
 Complimentary study~\cite{KiselevFeigelman} of superconductivity in STO due to exchange by
2TO phonons produced very reasonable results with a similar value $\lambda=1.1$ for the dimensionless coupling constant.
Our kinetic equation approach correctly reproduce temperature dependence of conductivity in various regions of moderate to high temperature. In particular, we obtained $\sigma(T) \propto T^{-3}$ behavior between about 100 and 200 Kelvin 
and slower dependence $\sigma(T) \propto T^{-3/2}$ at highest temperatures (see Fig.~\ref{alpha}), in agreement with experimental data~\cite{Behnia2020}.


The major difference between our approach and the theory part of Ref.~\cite{Behnia2020} is that we take into account 
energy-dependence $1/\tau_{TO}(p)$ which spoils simple proportionality between Seebeck coefficient and entropy per particle.
The same energy-dependence leads to non-zero Nernst coefficient which we calculate in the linear approximation over magnetic field.


We have used phenomenological notion of effective mass of electron, in spite of the fact that electron bands in STO are
rather anisotropic (apart from the very bottom of the first band). Partial justification for the use of such approximation
comes about since we consider high-temperature non-degenerate region, where electrons frequently change their energy
as well as momentum.  Independent check for the value of the effective mass about $10m_0$  we used comes from the
analysis of the plasma frequency measurements~\cite{Mass}.  Indeed, these data  for electron density
  $n=1.5\cdot 10^{20}$ cm$^{-3}$ show the drop of plasma frequency by the factor $\sim 1.5$ while temperature grows from
1 to 200 K. Using the relation $m(T)\propto 1/\omega_p^2(T)$ we conclude that effective mass grows by the factor $\sim 2.2$
in the same temperature range. Since low-temperature band mass is close to  $4m_0$ for this electron density, we find
$m_{eff}(T\sim 200) \approx 9m_0$, in good agreement with the value we used in our main analysis.


With the values obtained for  our fit parameters $\lambda$ and $m(T)$, we find average scattering 
rate $\langle 1/\tau_{TO}\rangle$ which is below, or at most, 20$\%$  higher, than the inverse
 Planckian time $k_B T/\hbar$, see inset to Fig.~\ref{fig:1}b. The  energy-dependent 
relaxation time $\tau_{TO}(p)$ never becomes less than 60$\% $ than Planckian time,  and is longer than 
it in the major part of the data shown in Fig.~\ref{fig:1}a.  Therefore we expect that the use of standard kinetic equation
we employed is possible for the problem studied, although corrections to it  may occur to be non-negligible at highest 
temperatures.

Recent paper~\cite{Yudson21} (which we became aware of after the first version of this paper was submitted) is devoted to
the $T$-dependence of conductivity in STO due to scattering by 2TO phonons. While the basic mechanism of electron scattering
considered in~\cite{Yudson21} is the same as in our paper, it is treated  in a quite different way.  Namely, scattering
by 2TO phonons is considered as  \textit{effectively elastic} electron scattering by "thermal disorder" produced by
slow TO phonons. In result, electron scattering rate and resistivity are found to scale as $T^2/E_0$ at low temperatures
$T \leq E_0$, where $E_0$ depends on the strength of the coupling to 2TO phonons (estimated as $\approx 200$ K in
Ref.~\cite{Yudson21}). In addition, scattering rate  they found is independent on the electron energy due to approximation
of zero TO phonon gap employed. Both these features of the low-$T$ results of Ref.~\cite{Yudson21} differ from our results
as well as from experimental observations; we believe that  their major problem is due to elastic approximation. 
However, at high temperature $T \geq E_0$ they obtained $\sigma \propto T^{-3/2}$, via non-trivial solution beyond quasiparticle approximation. This result is in qualitative agreement with our highest-T  results obtained within kinetic equation approach;
this coincidence seems to indicate that breakdown of quasiparticle description due to $\hbar/T > \tau_{TO}(p)$ does not lead to
substantial consequences for conductivity.
\section{Acknowledgments}

We thank Kamran Behnia and  Cl\'ement Collignon for  many useful discussions of experimental aspects of STO
and for providing us with the raw experimental data from \citep{Behnia2020};  we also thank Dirk van der Marel and
Igor Mazin for clarifications about the issue of effective electron mass in STO.
We are grateful to Anton Andreev, Denis Basko, Alexei Ioselevich, Dmitry Kiselev and Vladimir Yudson
 for their comments on the theory side.
This research  was supported by the Russian Science Foundation grant  \# 21-12-00104.

\appendix
\section{\label{app:collint} Collision integral}

 In order to treat the Hamiltonian \eqref{H1}
we must represent
the phonon operators in the second quantization form
\begin{align}
 & \hat{\boldsymbol{u}}(\boldsymbol{r})=\frac{1}{\sqrt{N}}\sum_{\boldsymbol{k},\lambda}\frac{\boldsymbol{e}^{\lambda}_{\boldsymbol{k}}}{\sqrt{2M\omega_{\boldsymbol{k}}}}\left(\hat{a}^{\lambda}_{\boldsymbol{k}}e^{-i\omega_{\boldsymbol{k}}t+i\boldsymbol{k}\boldsymbol{r}}+e.c\right)
\end{align}

where $\hat{a}^\lambda_{\boldsymbol{k}}$
is phonon annihilation operator, $M$
is the mass of the unit sell, $N$ is the number of cells and $\boldsymbol{e}^{\lambda}_{\boldsymbol{k}}$
is the $\lambda$ polarization vector. This rewrites the Hamiltonian in a form
\begin{align}
\hat{\mathcal{H}}_{TO} & =\frac{g}{2MN}\sum_{\boldsymbol{k_{1}},\boldsymbol{k_{2}},\lambda,\mu}\frac{\boldsymbol{e}^{\lambda}_{\boldsymbol{k_{1}}}\boldsymbol{e}^{\mu}_{\boldsymbol{k_{2}}}}{\sqrt{\omega_{\boldsymbol{k_{1}}}\omega_{\boldsymbol{k_{2}}}}}\left(\hat{\mathcal{F}}_{\boldsymbol{k_{1}k_{2}}}^{(1)\lambda\mu}+\hat{\mathcal{F}}_{\boldsymbol{k_{1}k_{2}}}^{(2)\lambda\mu}\right)\\
\hat{\mathcal{F}}_{\boldsymbol{k_{1}k_{2}}}^{(1)\lambda\mu} & =\left(\hat{a}^{\lambda}_{\boldsymbol{k_{1}}}\hat{a}^{\mu}_{\boldsymbol{k_{2}}}e^{i\left(\boldsymbol{k_{1}}+\boldsymbol{k_{2}}\right)\boldsymbol{r}}\right)e^{-i(\omega_{\boldsymbol{k_{1}}}+\omega_{\boldsymbol{k_{2}}})t}+e.c.\\
\hat{\mathcal{F}}_{\boldsymbol{k_{1}k_{2}}}^{(2)\lambda\mu} & =\left(\hat{a}_{\boldsymbol{k_{1}}}^{\lambda}\hat{a}_{\boldsymbol{k_{2}}}^{\dagger\mu}e^{i\left(\boldsymbol{k_{1}}-\boldsymbol{k_{2}}\right)\boldsymbol{r}}\right)e^{-i(\omega_{\boldsymbol{k_{1}}}-\omega_{\boldsymbol{k_{2}}})t}+e.c
\end{align}
For the scattering probability we will use Fermi's Golden rule.
Here, each term in the Hamiltonian gets a simple physical interpretation.
$\hat{\mathcal{F}}_{\boldsymbol{k_{1}k_{2}}}^{(1)}$ will stand for the processes
when two phonons are absorbed or emitted, and $\hat{\mathcal{F}}_{\boldsymbol{k_{1}k_{2}}}^{(2)}$
for the process of scattering on a phonon.

Now, we take into account $\langle\langle\hat{\psi}_{\boldsymbol{p}}^{\dagger}\hat{\psi}_{\boldsymbol{p}}\rangle\rangle=f_{\boldsymbol{p}}, \langle\langle\hat{\psi}_{\boldsymbol{p}}\hat{\psi}_{\boldsymbol{p}}^{\dagger}\rangle\rangle=1-f_{\boldsymbol{p}}, \langle\langle\hat{a}_{\boldsymbol{k}}^{\dagger}\hat{a}_{\boldsymbol{k}}\rangle\rangle=N_{\boldsymbol{k}}, \langle\langle\hat{a}_{\boldsymbol{k}}\hat{a}_{\boldsymbol{k}}^{\dagger}\rangle\rangle=N_{\boldsymbol{k}}+1$,
where $\hat{\psi}_{\boldsymbol{p}}^{\dagger}, \hat{\psi}_{\boldsymbol{p}}$
are electron creation and annihilation operators respectively, and
$f_{\boldsymbol{p}}$ and $N_{\boldsymbol{k}}$ are non-equilibrium
distributions for electrons and phonons respectively. The last step is to represent the electron
distribution function as follows $f_{\boldsymbol{p}}\approx n_{p}+\delta n_{\boldsymbol{p}}$ and take phonon distribution to be
almost equilibrium $N_{\boldsymbol{k}}\approx N_{k}$. 
Finally, the collision integral vanishes when all the distributions are taken to be the equilibrium ones.
 Therefore, saving only the terms linear with respect to $\delta n_{\boldsymbol{p}}$ we end up with \eqref{ITO}

\section{\label{sec:Tau} Parametrization of the collision integral}

The collision integral is of the form:
\begin{align}
I_{TO}\{\delta n_{\boldsymbol{p}}\}=\int_{\boldsymbol{k_{1}k_{2}}} f_1 (k_1,k_2,p) \delta n_{\textbf{p}} + f_2 (k_1,k_2,p) \delta n_{\textbf{p'}}
\end{align}
 where the functions $f_1, f_2$ were found in \eqref{ITO}. The momentum $p'$ is expressed via $\boldsymbol{k_{1}},$ $\boldsymbol{k_{2}},$ $\boldsymbol{p}$. Let us fix the function $\delta n_{\textbf{p}}$. After the integration over $\boldsymbol{k_{1}}$, $\boldsymbol{k_{2}}$ we will get a function of the momentum $p$: $I_{TO}\equiv I_{TO}(\textbf{p})$. Now, for that fixed $\delta n_{\textbf{p}}$ we can introduce a function:
\begin{align}
\tau_{TO}(p)=-\frac{\delta n_{\textbf{p}}}{I_{TO}(\textbf{p})}
\end{align}
As a result we will have an exact parametrization of our collision integral in a form:\begin{align}
I_{TO}(\textbf{p})=-\frac{\delta n_{\textbf{p}}}{\tau_{TO}(p)}
\end{align}
The function $\tau_{TO}(p)$ introduced in this way depends indeed on the function $\delta n_{\textbf{p}}$ itself. But the physical problem presumes an existence of some source of deviation from the Fermi distribution which results from a certain external sources (electric field, temperature gradient,  etc.), which means that in every specific problem  $\tau(p)$ can be determined.
It is not guaranteed  that the same $\tau(p)$ will be obtained for  different types of perturbation.

Our method looks apparently similar to the relaxation time approximation (RTA), however, in contrast to RTA, we do not make any approximations and use an exact parametrization of the collision integral. For sake of simplicity, we  call $\tau_{TO}(p)$ \textit{effective relaxation time.}

\section{\label{app:calc} Calculation of the relaxation time}
\subsection{\label{sec:calc1} The method of successive approximations}

\begin{figure*}[t]
{ \begin{centering}
\includegraphics[width=0.98\linewidth]{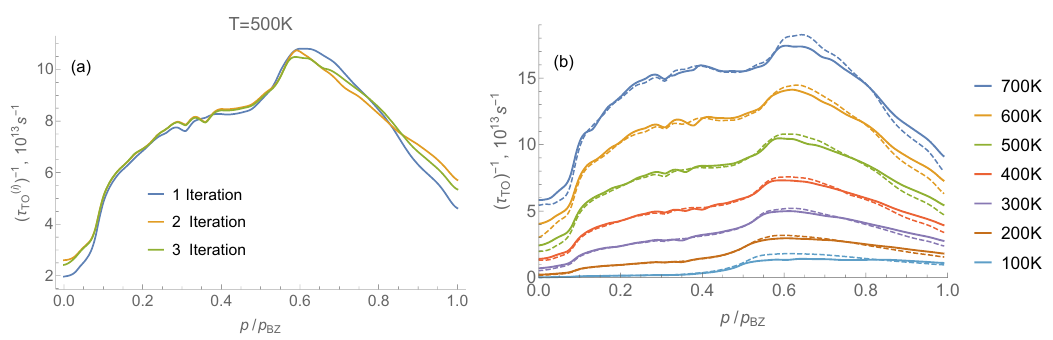}
\par\end{centering}}
\caption{\label{fig:MethodResultsApp} These plots represent the results of the implemented nethod of successive approximations. $p_{BZ}=\hbar \pi /a$ stands for the first Brillouin zone. \textbf{a)} The relaxation rate ${1}/{\tau_{TO}^{(i)}}$ for $i=1,2,3$ iterations for a fixed temperature $T=500$K \textbf{b)} The relaxation rates ${1}/{\tau_{TO}^{(i)}}$ for $i=1,3$ iterations for the considered range of temperatures}
\end{figure*} 

\begin{figure*}[t]
\begin{centering}
\includegraphics[width=0.9\textwidth]{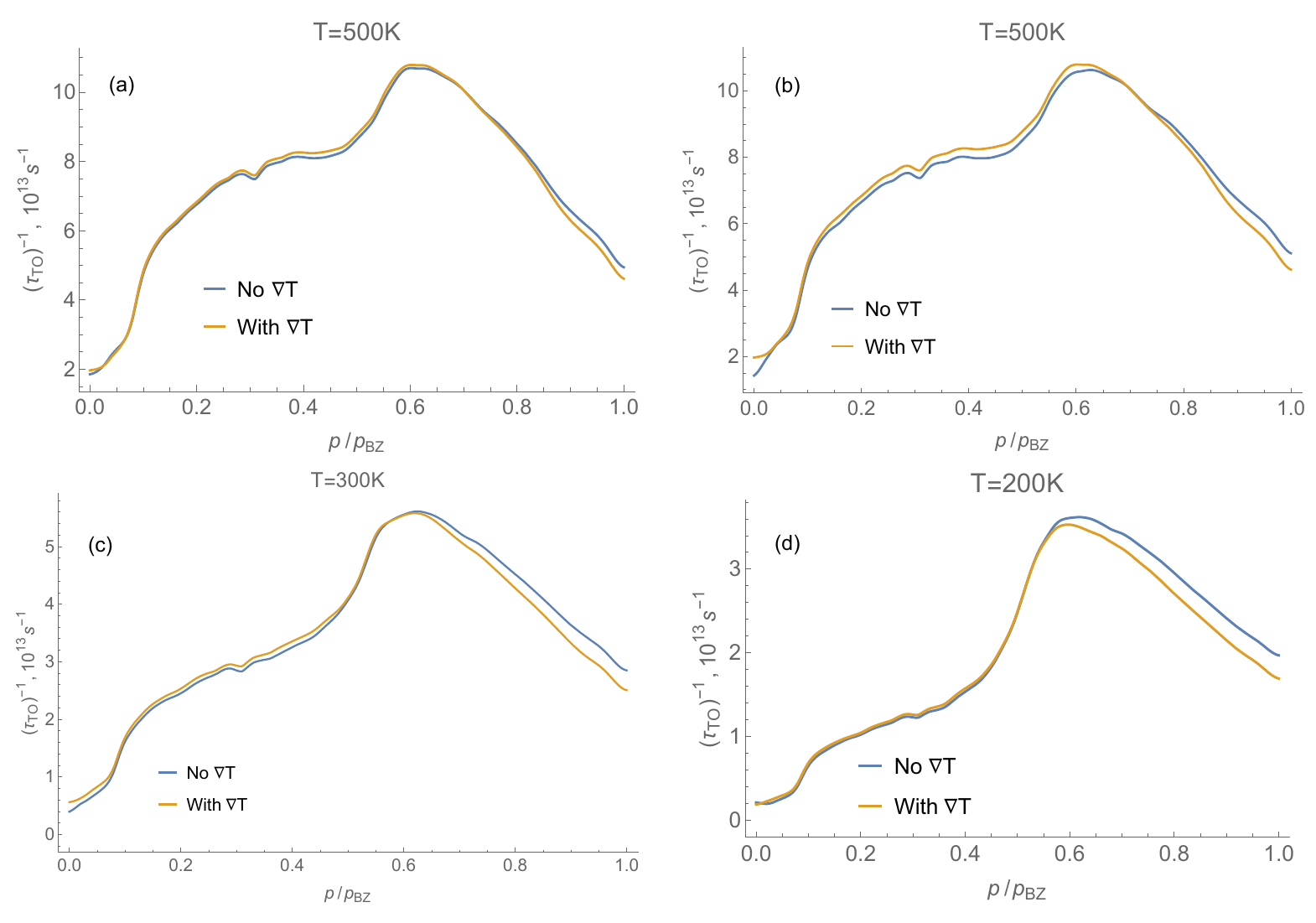}
\par\end{centering}
\caption{\label{fig:GradVSnoGrad} The results for the relaxation time with and without the temperature gradient considered for two values of  the Seebeck coefficient \textbf{a)} $\mathcal{S}= 700$ $\mu$V/K, \textbf{b)-d)} $\mathcal{S}= 300$ $\mu$V/K}
\end{figure*} 

In order to find the relaxation time we assume that the temperature gradient in the Boltzmann equation can be neglected. This assumption is checked in the Appendix~\ref{app:Results}. Therefore, we can rewrite the equation for $\delta n_{\textbf{p}}$ in the following form
\begin{align}
 & \delta n_{\boldsymbol{p}}\approx e\tau_{TO}(p)\cdot\left(\boldsymbol{v_{p}},\mathbf{E}\right)\frac{\partial n_{F}\left(\xi_{p}\right)}{\partial\xi_{p}}\label{eq:BeqN}\\
 & I_{TO}\{\delta n_{\boldsymbol{p}}\}\equiv-\frac{\delta n_{\boldsymbol{p}}}{\tau_{TO}(p)}\label{eq:RTAN}
\end{align}

 Now, let us introduce our modification of the method of successive
approximations for this case.\\
%
 \textit{ $0^{th}$ iteration:} We start from the simplest trial with an energy independent relaxation time
\begin{align}
 & \tau_{TO}^{(0)}=\text{const}\\
 & \delta n_{\boldsymbol{p}}^{(0)}\approx e\tau_{TO}^{(0)}\cdot\left(\boldsymbol{v_{p}},\mathbf{E}\right)\frac{\partial n_{F}\left(\xi_{p}\right)}{\partial\xi_{p}}\label{eq:n0}
\end{align}
%
 \textit{$1^{st}$ iteration:} We substitute \eqref{eq:n0} into \eqref{eq:RTAN}:
\begin{align}
 & I_{TO}^{(1)}\{\delta n_{\boldsymbol{p}}^{(0)}\}=\text{Integral}\{\delta n_{\boldsymbol{p}}^{(0)}\}\equiv-\frac{\delta n_{\boldsymbol{p}}^{(0)}}{\tau_{TO}^{(1)}(p)}\label{eq:t1}\\
 & \longrightarrow\tau_{TO}^{(1)}(p)=-\frac{\delta n_{\boldsymbol{p}}^{(0)}}{\text{Integral}\{\delta n_{\boldsymbol{p}}^{(0)}\}}
\end{align}
where the word ``Integral'' means that we numerically evaluate the
integral in \eqref{ITO} using $\delta n_{\boldsymbol{p}}^{(0)}$
defined with \eqref{eq:n0}. Here, we know $\delta n_{\boldsymbol{p}}^{(0)}$
as a function of $\boldsymbol{p}$ to within the constant factor $\tau_{TO}^{(0)}$.
The key part is that $I_{TO}^{(1)}\{\delta n_{\boldsymbol{p}}^{(0)}\}$
is linear with respect to $\delta n_{\boldsymbol{p}}^{(0)}$, which means that the value of $\tau_{TO}^{(0)}$
cancels out from Eq.\eqref{eq:t1}.  
 This allows us to extract $\tau_{TO}^{(1)}$. Using it, we find modified distribution function
\begin{align}
 & \delta n_{\boldsymbol{p}}^{(1)}\approx e\tau_{TO}^{(1)}(p)\cdot\left(\boldsymbol{v_{p}},\mathbf{E}\right)\frac{\partial n_{F}\left(\xi_{p}\right)}{\partial\xi_{p}}\label{eq:n1}
\end{align}

 \textit{$2^{nd}$ iteration:} Similarly,
\begin{align}
 & I_{TO}^{(2)}\{\delta n_{\boldsymbol{p}}^{(1)}\}=\text{Integral}\{\delta n_{\boldsymbol{p}}^{(1)}\}\equiv-\frac{\delta n_{\boldsymbol{p}}^{(1)}}{\tau_{TO}^{(2)}(p)}
\end{align}
Here, we already know $\delta n_{\boldsymbol{p}}^{(1)}$, so no problems
in finding $\tau_{TO}^{(2)}$. And then 
\begin{align}
 & \delta n_{\boldsymbol{p}}^{(2)}\approx e\tau_{TO}^{(2)}(p)\cdot\left(\boldsymbol{v_{p}},\mathbf{E}\right)\frac{\partial n_{F}\left(\xi_{p}\right)}{\partial\xi_{p}}\label{eq:n1-1}
\end{align}

 \textit{$n^{th}$ iteration:}
\begin{align}
 & I_{TO}^{(n)}\{\delta n_{\boldsymbol{p}}^{(n-1)}\}=\text{Integral}\{\delta n_{\boldsymbol{p}}^{(n-1)}\}\equiv-\frac{\delta n_{\boldsymbol{p}}^{(n-1)}}{\tau_{TO}^{(n)}(p)}\label{eq:tn}\\
 & \delta n_{\boldsymbol{p}}^{(n)}\approx e\tau_{TO}^{(n)}(p)\cdot\left(\boldsymbol{v_{p}},\mathbf{E}\right)\frac{\partial n_{F}\left(\xi_{p}\right)}{\partial\xi_{p}}\label{eq:nn}
\end{align}\\
The convergence of the presented method can not be guaranteed. However, we are working with a strictly positive function $\tau_{TO}(p)>0$; hence it has a strictly positive average value which can be considered as the $\tau_{TO}^{(0)}$. So, we
can think of $\tau_{TO}(p)$ as $\tau_{TO}(p)=\overline{\tau_{TO}}+\Delta\tau_{TO}(p)$
and if $\Delta\tau_{TO}(p)$ is not much greater compared to $\overline{\tau_{TO}}$, the method should converge.

Now, let us prove that if the numerical evaluation shows that the method converges
$\big(\lim\limits_{n\to\infty}\tau_{TO}^{(n)}=\tau_{TO}^{(\infty)};$ $\quad\lim\limits_{n\to\infty}\delta n_{\boldsymbol{p}}^{(n)}=\delta n_{\boldsymbol{p}}^{(\infty)}\big)$,
we can guarantee that the obtained result is the solution for \eqref{eq:BeqN}
and \eqref{eq:RTAN}. The reason for this is that if we formally take a limit
$n\to\infty$ in \eqref{eq:tn} and \eqref{eq:nn}, we end up with:
\begin{align}
 & \delta n_{\boldsymbol{p}}^{(\infty)}\approx e\tau_{TO}^{(\infty)}(p)\cdot\left(\boldsymbol{v_{p}},\mathbf{E}\right)\frac{\partial n_{F}\left(\xi_{p}\right)}{\partial\xi_{p}}\label{eq:Beq-1}\\
 & I_{TO}\{\delta n_{\boldsymbol{p}}^{(\infty)}\}=\text{Integral}\{\delta n_{\boldsymbol{p}}^{(\infty)}\}\equiv-\frac{\delta n_{\boldsymbol{p}}^{(\infty)}}{\tau_{TO}^{(\infty)}}\label{eq:RTA-1}
\end{align}
The  above pair of equations is fully equivalent to Eqs. \eqref{eq:BeqN} and \eqref{eq:RTAN}, which concludes our proof.

\subsection{\label{app:Results} Results for $\tau_{TO}(p)$ and the role of  temperature gradient}

The results of the implemented method are presented in Fig. \ref{fig:MethodResultsApp}. It is clear that the method converges very decently, as the difference between the third approximation and the second one is already quite small. Therefore, it is sufficient to consider $\tau_{TO}^{(3)}(p)$ to be our final approximation for the relaxation time. 

As it is clear from  Eq.\eqref{eq:BeqN}, we neglected the temperature gradient while developing the method of successive approximations. Let us now check whether this approximation was appropriate. In order to do that we assume the absence of the charge current $\mathbf{J}_{\mathbf{E}}=0$ which gives a correlation between the electric field and the temperature gradient $\nabla T= E/\mathcal{S}$. We consider several temperatures and two values of the Seebeck coefficient. The Fig.~\ref{fig:GradVSnoGrad} shows the comparison between the relaxation times computed with and without the temperature gradient.



\begin{thebibliography}{99}

\bibitem{Behnia2020} Cl\'ement Collignon, Phillipe Bourges, Beno\^it Fauqu\'e, and Kamran Behnia, "Heavy non-degenerate electrons in doped strontium titanate", Phys. Rev. X \textbf{10},  031025 (2020)

\bibitem{Spinelli2010} A. Spinelli, M. A. Torija, C. Liu, C. Jan, and C. Leighton,
"Electronic transport in doped SrTiO3: Conduction mechanisms and potential applications",
Phys. Rev.  B \textbf{81}, 155110 (2010).


\bibitem{review1}  Maria N.Gastiasoro, Jonathan Ruhman and Rafael M.Fernandes, "Superconductivity in dilute SrTiO: a review",
  Annals of Physics, \textbf{417}, 168107  (2020)

\bibitem{review2}  Cl\'ement Collignon, Xiao Lin, Carl Willem Rischau, Beno\^it Fauqu\'e, and Kamran Behnia,   "Metallicity and Superconductivity in Doped Strontium Titanate",   Annual Review of Condensed Matter Physics,  \textbf{10},  25 (2019)

\bibitem{Behnia2017} Lin, X., Rischau, C.W., Buchauer, L. et al.  "Metallicity without quasi-particles in room-temperature strontium titanate", Nature Quant. Mater. \textbf{2},  41 (2017). https://doi.org/10.1038/s41535-017-0044-5

\bibitem{Verma} 
A. Verma, A. P. Kajdos, T. A. Cain, S. Stemmer, and D. Jena,
"Intrinsic Mobility Limiting Mechanisms in Lanthanum-Doped Strontium Titanate",
  Phys. Rev. Lett. \textbf{112}, 216601 (2014)

\bibitem{Cain} T. A. Cain, A. P. Kajdos, and S. Stemmer, "La-doped SrTiO3 films with large cryogenic
thermoelectric power factors",  Appl. Phys. Lett. \textbf{102}, 182101 (2013).



\bibitem{YamadaShirane} Yamada, Y. \& Shirane, G. "Neutron scattering
and nature of the soft optical phonon in SrTiO3". J. Phys. Soc. Jpn.
\textbf{26}, 396\textendash 403 (1969).

\bibitem{Levanyuk1}  
 Yu.N.Epifanov, A.P. Levanyuk and G.M. Levanyuk, Fiz.Tverd.Tela, \textbf{23}, 690 (4981).


\bibitem{Levanyuk2}  Yu. N. Epifanov , A. P. Levanyuk and G. M. Levanyuk 
"Interaction of carriers with to-phonons and electrical conductivity of ferroelectrics", 
Ferroelectrics, \textbf{35}, 199-202 (1981).

\bibitem{old1} O. N. Tufte and P. W. Chapman, "Electron mobility in semiconducting STO",
 Phys. Rev. \textbf{155}, 796 (1967).

\bibitem{old2} H. P. R. Frederikse and W. R. Hosler, "Hall mobility in SrTiO3",
Phys. Rev. \textbf{161}, 822 (1967).

\bibitem{Yudson21} A. Kumar, V. I. Yudson and D. L. Maslov,
 Phys. Rev. Lett. \textbf{126}, 076601 (2021).


\bibitem{Dirk2011}  D. van der Marel, J. L. M. van Mechelen, and I. I. Mazin,
"Common Fermi-liquid origin of T$^2$ resistivity and superconductivity in n-type SrTiO3",
Phys. Rev. B \textbf{84}, 205111 (2011).


\bibitem{numeric1} B. Himmetoglu, A. Janotti, H. Peelaers, A. Alkauskas,
                and C. G. Van de Walle, "First-principles study of the mobility of SrTiO3",
Phys. Rev. B \textbf{90}, 241204(R) (2014).

\bibitem{numeric2}  J.-J. Zhou, O. Hellman, M. Bernardi, "Electron-Phonon Scattering in the Presence of Soft Modes and Electron Mobility in SrTiO3 Perovskite from First Principles",
	Phys. Rev. Lett. \textbf{121}, 226603 (2018)

\bibitem{numeric3} J.-J. Zhou and M. Bernardi, "Predicting charge transport in the presence of polarons:
The beyond-quasiparticle regime in SrTiO3", Phys. Rev. Research \textbf{1}, 033138 (2019).



\bibitem{soundvel} R. O. Bell and G. Rupprecht, Phys. Rev. 129, 90 (1963)

\bibitem{defpot} A.N.Morozovska, E.A.Eliseev, G.S.Svechnikov and S.V.Kalinin, 
Phys. Rev. B \textbf{84}, 045402 (2011)

\bibitem{acoustic} C. Hamaguchi, Basic Semiconductor Physics (Springer,
New York, 2010), 2nd ed.

\bibitem{Theory} W.-R. Lee, A. M. Finkel\textquoteright stein,
K. Michaeli, and G. Schwiete,
Phys. Rev. Research \textbf{2}, 013148 (2020).


\bibitem{Behnia2014} X. Lin, G. Bridoux, A. Gourgout, G. Seyfarth,
S. Kr\"amer, M. Nardone, B. Fauqu\'e, and K. Behnia,
"Critical Doping for the Onset of a Two-Band Superconducting Ground State in SrTiO$_{3-\delta}$",
Phys. Rev. Lett. \textbf{112}, 207002 (2014).

\bibitem{Baurle}B\"auerle, D., Wagner, D., W\"ohlecke, M., Dorner, B.
\& Kraxenberger, H. Soft modes in semiconducting SrTiO3: II. The ferroelectric
mode. Z. Physik B\textemdash Condensed Matter. 38, 335\textendash 339
(1980).


\bibitem{Dirk2019}  D. van der Marel, F. Barantani, and C. W. Rischau, Phys. Rev. Res. \textbf{1}, 013003 (2019).

\bibitem{KiselevFeigelman} D. Kiselev and M. Feigel'man,
 "Theory of superconductivity due to Ngai’s mechanism in lightly doped SrTiO3",
arxiv:2106.09530


\bibitem{T^2} Lin, X., Fauqu\'e, B. \& Behnia, K. Scalable T$^2$ resistivity
in a small single-component Fermi surface. Science 349, 945 (2015).

\bibitem{Svel} W. Rehwald, Solid State Communications 8, 607 (1970).

\bibitem{KamranBook} Kamran Behnia, \textit{Fundamentals of Thermoelectricity}, Oxford University Press, 2019.



\bibitem{Behnia2013} X. Lin, Z. Zhu, B. Fauqu\'e, and K. Behnia, "Fermi Surface of the Most Dilute Superconductor",
Phys. Rev. X \textbf{3}, 021002 (2013).


\bibitem{Mass}  J. L. M. van Mechelen, D. van der Marel, C. Grimaldi, A. B. Kuzmenko, N. P. Armitage, N. Reyren,
H. Hagemann, and I. I. Mazin, "Electron-Phonon Interaction and Charge Carrier Mass Enhancement in SrTiO3",
Phys. Rev. Lett. \textbf{100}, 226403 (2008).





\end{thebibliography}
\end{document}